\documentclass[a4paper,11pt]{article}
\usepackage{graphicx}
\usepackage{xcolor}
\usepackage{amsmath,amsfonts,amssymb,amstext,graphicx}
\usepackage{cancel}
\usepackage{empheq}
\usepackage{float}
\usepackage{subcaption}
\usepackage{authblk}
\usepackage{blindtext}
\usepackage{cases}
\usepackage{epsfig}%
\usepackage[bottom]{footmisc}

\newcommand{\dv}[2]{\frac{d#1}{d#2}}

\usepackage{epsfig}%
\usepackage[bottom]{footmisc}
\newcommand{\lsim}{\raisebox{-0.13cm}{~\shortstack{$<$ \\[-0.07cm]
      $\sim$}}~}

\begin{document}
\date{}

\title{\centerline \bf Modified scaling in $k$-essence model in interacting dark energy - dark matter scenario}
\bigskip
\author[1]{Anirban Chatterjee \thanks{Corresponding Author: anirbanc@iitk.ac.in}}
\author[2]{Biswajit Jana\thanks{vijnanachaitanya2020@gmail.com}}
\author[2]{Abhijit Bandyopadhyay\thanks{abhijit.phy@gm.rkmvu.ac.in}}
\normalsize
\affil[1]{Indian Institute of Technology Kanpur, Kanpur 208016, India}
\affil[2]{Ramakrishna Mission Vivekananda Educational and Research Institute, 
Belur Math, Howrah 711202}
\date{\today}
\maketitle

\begin{abstract} 
It has been shown by  \textit{Scherrer and Putter et.al}. in  \cite{Scherrer:2004au,dePutter:2007ny} that, 
when dynamics of dark energy is driven by a homogeneous
$k-$essence scalar field $\phi$, with a Lagrangian of the form $L = V_0F(X)$
with a constant potential $V_0$ and
$X = \frac{1}{2}\nabla^\mu\phi \nabla_\mu\phi = \frac{1}{2}\dot{\phi}^2$,
one obtains a scaling relation $X(dF/dX)^2 = Ca^{-6}$ , where $C$ is a constant and $a$
is the FRW scale factor of the universe. 
The separate energy conservation in
the dark energy sector and the constancy of  $k-$essence potential are 
instrumental in obtaining such  a scaling. 
In this paper we have shown that, even when considering  time-dependent 
interactions between  
dark energy and dark matter,
the constancy  of $k-$essence potential may lead to a modified form of  
scaling. We have obtained such a scaling relation for a particular class of parametrisation 
of the source term occurring in the continuity equation of dark energy and dark matter
in the interacting scenario. We used inputs from the  JLA  analysis of 
luminosity distance and redshift data from Supernova Ia observations,
to obtain the modified form of the scaling.
\end{abstract}

\section{Introduction}
\label{seci}
 Several cosmological observations and surveys revealed  
  that our universe is presently undergoing  an accelerated phase of  
  expansion and  a transition from decelerated phase  to this accelerated phase happened
  during the late time phase of cosmic evolution.  
  The luminosity distance and redshift measurements of type Ia Supernova   
  \cite{Riess:1998cb,Perlmutter:1998np,Riess:2006fw} are instrumental in establishing this fact. 
  Independent observations like Baryon Acoustic Oscillation \cite{Eisenstein:2005su, Percival:2006gs}, 
  Cosmic Microwave Background Radiation \cite{Gawiser:2000az} and Observed Hubble Data (OHD) \cite{OHD} 
  also reinforce this conclusion. 
  The cause of   acceleration of the present-day cosmic expansion still remains a mystery  
  and has been presented in  literature  
  \cite{Riess:1998cb,Perlmutter:1998np,Riess:2006fw} as an effect due to presence 
  of an unclustered form of energy with   negative pressure - dubbed `Dark Energy' (DE).
  On the other hand, observed rotation curves of spiral galaxies \cite{Sofue:2000jx}, observation of 
  gravitational lensing \cite{Bartelmann:1999yn}, bullet  and other colliding clusters provide
  evidences for existence 
  of non-luminous matter in the universe, called `Dark Matter' (DM), which reveals its presence 
  only through gravitational effects detected in above mentioned observations.
  Results from satellite borne experiments like WMAP \cite{Hinshaw:2008kr} and Planck \cite{Ade:2013zuv} have 
  estimated that DE and DM together contribute $\sim 96\%$ of total energy density of the 
  present-day universe, with  $\sim$69$\%$  and $\sim$27$\%$ as their respective shares.
  The rest  4$\%$ contribution comes from radiation and baryonic matter.   
 Though a physical theory of  dark energy is still lacking, there exist diverse theoretical approaches aiming 
 construction of models for DE  leading to present-day cosmic acceleration.
 These include the $\Lambda$-CDM model, where $\Lambda$ refers to cosmological constant and `CDM' corresponds 
 to cold (non-relativistic) dark matter. This model, though fits well with the present-day 
 cosmological  observations,
 is associated with the problems of cosmological coincidence 
 \cite{Zlatev:1998tr} and fine tuning  \cite{Martin:2012bt}
 which motivate construction  of  alternative DE models. One approach of constructing
 such models, called modified gravity models \cite{fr1}, 
 involves modification of Einstein tensor in the geometric part of 
 the Einstein field equations. Another class of models treats DE to be 
 driven by (scalar) fields with 
 suitably chosen field theoretic Lagrangians contributing to energy-momentum tensor in Einstein 
 equations.
 The second kind of models, as widely discussed in literature, includes the Quintessence 
\cite{Peccei:1987mm,Ford:1987de,Peebles:2002gy,Nishioka:1992sg, Ferreira:1997au,Ferreira:1997hj,Caldwell:1997ii,Carroll:1998zi,Copeland:1997et} and $k$-essence models  \cite{Fang:2014qga,ArmendarizPicon:1999rj,ArmendarizPicon:2000ah,ArmendarizPicon:2000dh,ArmendarizPicon:2005nz,Chiba:1999ka,ArkaniHamed:2003uy,Caldwell:1999ew}.  
 A sub-class of such models involves consideration of interaction between DE and DM 
 \cite{Szydlowski, calogero, calogero1, calogero2, Haba:2016swv,Ber1,Ber2,Band1,Band2,Band3} to explain the 
 present-day observed features of cosmic expansion. We have considered such a model  
 in the context of this paper.\\

 To investigate the interacting scenario of DE and DM, 
we neglect the radiation and baryonic matter contribution to the total energy
density of universe during its late time phase of cosmic evolution 
due to their small share ($\sim 4\%$) in the present-day
energy-content as estimated from WMAP \cite{Hinshaw:2008kr} and Planck \cite{Ade:2013zuv} observations.
In this paper, we consider DE to be represented by a homogeneous scalar field $\phi = \phi(t)$
whose dynamics is driven by a purely kinetic ($k$-essence) Lagrangian with a constant potential. Purely
kinetic scalar field models have been widely discussed in \cite{Scherrer:2004au,ArmendarizPicon:1999rj,ArmendarizPicon:2000ah,Chiba:1999ka,Garriga:1999vw,Armendariz-Picon:2000nqq,Chimento:2003zf,Chimento:2003ta}
and references therein. The DM component of the universe, on the other hand, is chosen to be
a non-relativistic pressure-less fluid (dust). We consider the DE-DM interaction  
 to be time-dependent and introduce it through a source term $Q(t)$ 
in the  non-conserving continuity equations for DM and DE
(see  Eqs.\ (\ref{eq:contdm}) and  (\ref{eq:contde}) in Sec.\ \ref{seciii}).
The function $Q(t)$ provides a measure of instantaneous rate of
energy transfer between DE and DM components of the universe.\\

We consider  spacetime geometry of the expanding universe, at large scales, to be 
isotropic and homogeneous and flat which is described by  the 
Friedmann - Robertson - Walker (FRW) metric characterised by a time 
dependent  scale factor $a(t)$ and zero curvature constant. 
The evolution of such a universe with its content modelled as
an ideal fluid (characterised by its energy density $\rho(t)$ and pressure $p(t)$) 
is governed by the Friedmann equations which connect $a(t)$ and its derivatives
with $\rho(t)$ and $p(t)$. 
Using the luminosity distance \textit{vs.} redshift measurements in Supernova Ia (SNe Ia) observations,
the temporal profile of the scale factor, the Hubble parameter $H = \dot{a}/a$ may be extracted.
From this knowledge, the time profile of  total energy density and pressure of the universe
may as well be computed, exploiting the Friedmann equations.\\

When we consider DM and DE to be non-interacting,
both  the components of  total dark fluid  
separately satisfy respective continuity equations (Eqs.\ (\ref{eq:contdm}) and
(\ref{eq:contde}) with $Q(t)=0$) implying separate 
energy conservation in each individual sector.   
In this non-interacting DE-DM scenario, if the
dynamics of DE is considered to be driven by a  
scalar field $\phi$ governed by a $k$-essence Lagrangian of form $L = V_0F(X)$
with a constant potential $V_0$
and  $X = \frac{1}{2}\nabla^\mu\phi \nabla_\mu\phi = \frac{1}{2}\dot{\phi}^2$,
one obtains a   scaling   of the form: $X(dF/dX)^2 = Ca^{-6}$ 
($C$ is a constant) \cite{Scherrer:2004au,dePutter:2007ny}. Separate energy conservation in
the DE sector and the constancy of  $k-$essence potential are instrumental 
in obtaining such   a scaling. This scaling relation  connects the scale factor 
$a$ and $k$-essence scalar field $\phi$  
through its time derivatives appearing in $X$ and $dF/dX$. \\

In this paper we have shown that, even in presence of   time-dependent interactions
between DE and DM ($Q(t) \neq 0$) implying continual exchange of
energy between the two sectors,
the constancy of $k-$essence potential  may lead  to a modified form of  
scaling. We have obtained such a scaling,
by parametrising the dependence of source term $Q(t)$ on $a(t)$ 
as a power law: $Q(t)=Q_0 [a(t)]^k$, where $k$ is a constant and $Q_0$
is the value of $Q(t)$ at present epoch ($a(t)$ is normalised to 1 at present epoch). 
In obtaining the form of scaling relation with above parametrisation of
DE-DM interactions, we have taken into consideration the observed feature of temporal
behaviour of the FRW scale factor $a(t)$, probed in the measurement of luminosity distance and
redshift of SNe Ia events. The observational ingredient enters into the 
explicit form of scaling relation at different levels of its derivation, 
through various constants which encode in them features of the observational data.
The obtained scaling depends on $k$ and $Q_0$, which are the two
parameters of the
model. For computational convenience,
in stead of $Q_0$, 
we used a dimensionless parameter $\beta_0 \equiv Q_0/(\rho_{\rm de}^0 + \rho_{\rm dm}^0)$ 
all through the work.
We also find the region of the corresponding $k - \beta_0$ 
parameter space that is  allowed from SNe Ia data,
in the context of interacting DE-DM model
considered here.  \\

The paper is organised as follows. In Sec.\ \ref{secii}, we  briefly discussed Joint Light curve Analysis (JLA)
of Supernova Ia data using which we obtain  the temporal behaviour of scale factor and its 
derivatives.  
 In Sec.\ \ref{seciia}, we gave a 
brief outline of the  framework  of   $k$-essence model of dark energy
and discussed its aspects for a constant potential scenario, which is relevant in the context
of the paper.
In Sec.\ \ref{seciii}, we presented the theoretical framework of   
interacting DE-DM scenario   in FRW universe. We  obtained temporal behaviour 
of equation of state parameter ($\omega = p/\rho$) of the total dark fluid and
 the total energy density $\rho$ of the universe 
using obtained time profile of the scale factor and its derivatives obtained in Sec.\ \ref{secii}.
In the context of this interacting 
DE-DM scenario, temporal behaviour of energy density of individual DE and DM  components
($\rho_{\rm de}$ and $\rho_{\rm dm}$) have also been obtained in terms of model-parameters $k$ and $\beta_0$.
In Sec.\ \ref{seciv} we have shown how we derived the corresponding modified scaling relation
  for $k-$essence model of DE with a constant
potential. The sensitivity of the modified scaling relation on the parameters $k$ and $\beta_0$
are also graphically
represented for different chosen benchmark values of the parameters within their allowed region.
We summarised the results in the concluding  Sec.\ \ref{secv}.

\section{Cosmological parameters from analysis of SNe Ia data}
\label{secii} 
\begin{figure}[H]	
\centering
\includegraphics[width=0.7\linewidth]{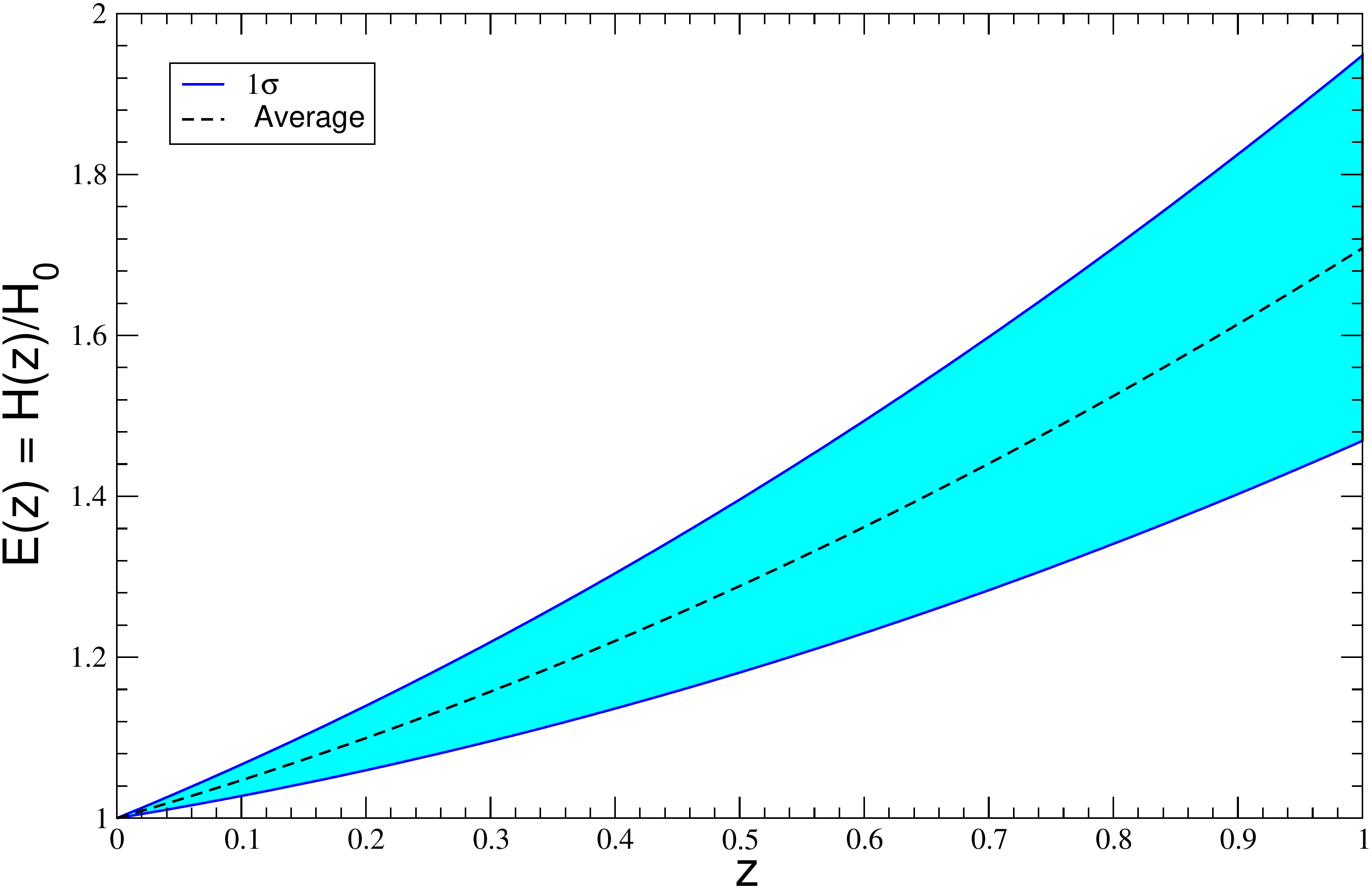} 
\caption{Plots of $E$ vs $z$ (Dashed line correspond to the average value of $E(z)$ over
the 1$\sigma$ range of $E(z)$) depicted by shaded region.}
\label{fig:EZ}
\end{figure}

As discussed in Sec.\ \ref{seci} the luminosity distance and redshift measurements
of type Ia Supernova is the key observational ingredient in establishing
the transition from decelerated to an accelerated phase of cosmic expansion
during its late time phase of evolution. We have used the observational
data as input, in estimating the behaviour of modified scaling,
on the parameters $k$, $\beta_0$ which parametrises the 
time dependent interaction $Q(t)$ between DE and DM.
In this section we describe how we extract the relevant cosmological parameters
from the SNe Ia data to use them as direct inputs into this estimation.  
There exist different compilations of the SNe Ia data corresponding
to  supernova surveys in different redshift region using diverse probes and measurements. 
Small redshift $(z > 0.1)$ projects comprise  Harvard-Smithsonian Center for Astrophysics survey (cFa) \cite{ref:Hicken}, the Carnegie Supernova Project (CSP)(\cite{ref:Contreras},\cite{ref:Folatelli},\cite{ref:Stritzinger}), the Lick Observatory Supernova Search (LOSS)  \cite{ref:Ganeshalingam}  and the Nearby Supernova Factory (SNF) \cite{ref:Aldering}.
SDSS-II supernova surveys (\cite{ref:Frieman},\cite{ref:Kessler},\cite{ref:Sollerman},\cite{ref:Lampeitl},\cite{ref:Campbell}) are mainly focused on the redshift region of ($0.05<z<0.4$). Programmes like Supernova Legacy Survey (SNLS) 
(\cite{ref:Astier},\cite{ref:Sullivan}) the ESSENCE project \cite{ref:Wood-Vasey}, the 
Pan-STARRS survey (\cite{ref:Tonry},\cite{ref:Scolnic},\cite{ref:Rest}) correspond
to high redshift regime.
More than one thousand SNe Ia events have been discovered through all surveys. However, 
in the range between $z \sim 0.01$ and $z \sim 0.7$, luminosity distances have been
measured with a very high  statistical precision.
`Joint Light-curve Analysis  (JLA) data' (\cite{ref:Scolnic},\cite{ref:Conley},\cite{ref:Suzuki})   contains a total of 740 SNe Ia events from
full three years of SDSS survey, first three seasons of
the five-year SNLS survey and 14 very high redshift $0.7<z<1.4$ SNe Ia from space-based observations with the HST  \cite{ref:Riess}.
For the present work, we consider this data set for analysis 
where the different systematic uncertainties are taken care of 
by compiling the data with flux-averaging 
technique whose technical details 
have been comprehensively 
described in \cite{jla,wang:jcap,wang:prd}. We briefly outline the methodology
of analysis here. The   $\chi^2$ function corresponding to JLA data is defined as
\begin{eqnarray}
\chi^2 = \sum_{i,j=1}^{740}(\mu_{\rm obs}^{(i)}  - \mu_{\rm th}^{(i)}) 
(\sigma^{-1})_{ij}
(\mu_{\rm obs}^{(j)}  - \mu_{\rm th}^{(j)} ) \,,
\label{eq:jla1}
\end{eqnarray}
where $\mu_{\rm obs}^{(i)}$ and $\mu_{\rm th}^{(i)}$ respectively
denote the observed value and theoretical expression for
distance modulus  at red-shift $z_i$. In a flat FRW spacetime,
$\mu_{\rm th}^{(i)}$ is given by
\begin{eqnarray}
\mu_{\rm th}^{(i)} &=& 
5\log_{10} [d_L(z_{\rm hel},z_{\rm CMB})/{\rm Mpc})] + 25
\label{eq:rnew1}
\end{eqnarray}
where  $d_L$ is the luminosity distance it's given by
\begin{eqnarray}
d_L(z_{\rm hel},z_{\rm CMB}) &=& 
(1 + z_{\rm hel}) r(z_{CMB}) \quad \mbox{with} \quad r(z) = cH_0^{-1} \int_0^z \frac{dz^\prime}{E(z^\prime)}\,,
\label{eq:rnew2}
\end{eqnarray}
and $z_{\rm CMB}$ and $z_{\rm hel}$ respectively 
refer to CMB rest frame and heliocentric frame
value of SNe IA redshifts. The value of Hubble parameter at present
epoch is denoted by $H_0$. The observed distance modulus  $\mu_{\rm obs}^{(i)}$
is expressed in terms of  the observed peak magnitude 
$m_B^\star$, the time stretching parameter of the light-curve 
$X_1$ and supernova color at maximum brightness, $C$ as
\begin{eqnarray}
 \mu_{\rm obs}^{(i)} &=& m_B^\star(z_i) - M_B + \alpha X_1(z_i) - \beta C(z_i)
\label{eq:rnew3}
\end{eqnarray}
$\alpha$, $\beta$ being the nuisance parameters and 
the absolute magnitude  $M_B$ is kept fixed at the value -19. $\sigma_{ij}$ 
in Eq.\ (\ref{eq:jla1}) is the covariant matrix  as given in 
Eq.\ (2.16) of \cite{wang:jcap}. 
Wang in \cite{wang:jcap} proposed a flux averaging technique  
to reduce the effect of systematic uncertainties involved in the covariant matrix
owing to weak lensing of SNe Ia data.  We take the result of
redshift($z$)-dependence of the function 
$E(z) = H(z)/H_0$ (corresponding to a zero red-shift cut-off $z=0$) 
obtained in \cite{wang:jcap} 
from the $\chi^2$-marginalisation with respect to $(M_B, \alpha, \beta)$.
The 1$\sigma$ range of $E(z)$ resulting from the analysis
is shown in Fig.\ \ref{fig:EZ}. The mean 
of $E(z)$ values for every $z$ in this $1\sigma$ range
 is shown by the dashed line in Fig .\ \ref{fig:EZ}.
The temporal behaviour of   relevant cosmological quantities
are obtained using this  mean $E(z)$ vs $z$
curve as benchmark.

\begin{figure}[H]	
\centering
\includegraphics[width=0.7\linewidth]{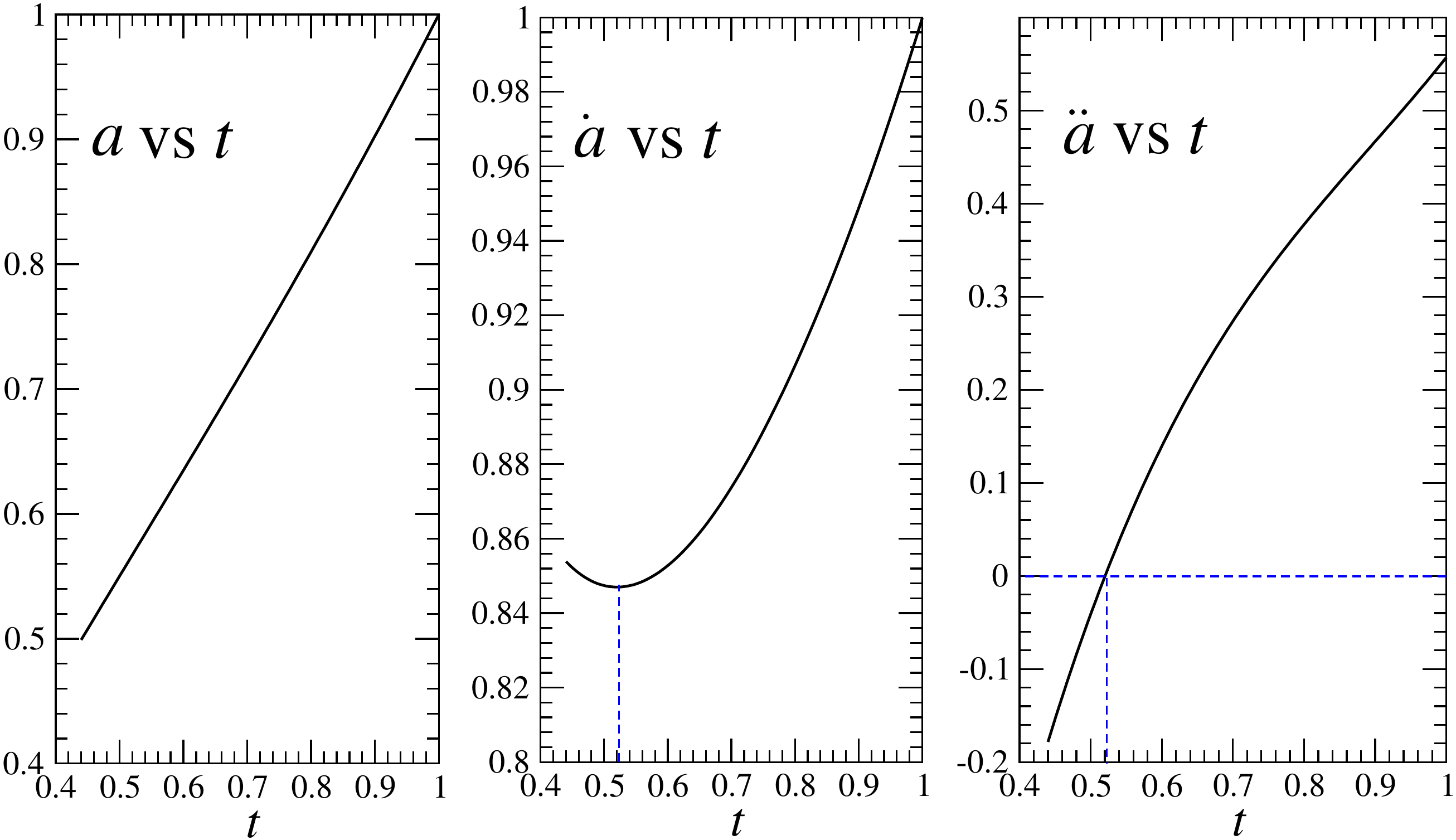} 
\caption{Plot of $a$ (left panel), $\dot{a}$ (middle panel) and $\ddot{a}$ (right panel) vs $t$
 corresponding to the central line of Fig.\ \ref{fig:EZ}.}
\label{fig:avst}
\end{figure}

The scale factor $a$, which we have taken
to be normalised to $a=1$ at present epoch, is related to redshift $z$ by the relation 
$1/a = 1+z$. Using  $H=\dot{a}/a$ we can write
\begin{eqnarray}
dt &=& - \frac{dz}{(1+z)H(z)} = - \frac{dz}{(1+z)H_0E(z)}
\label{eq:jla2}
\end{eqnarray}
which on integration gives
\begin{eqnarray}
\frac{t(z)}{t_0} &=& 1 
- \frac{1}{H_0t_0}\int_z^0 \frac{dz^\prime}{(1+z^\prime)E(z^\prime)}
\label{eq:jla3}
\end{eqnarray}
where $t_0$ denotes present epoch. Using $E(z)$ vs $z$ profile as shown in Fig.\ \ref{fig:EZ} 
we perform the above integration numerically  
to obtain $z$ dependence of $t(z)$. Using Eq.\ (\ref{eq:jla3}) and the relation $1/a = 1+z$,
simultaneous values of $a$ and $t$ at any given redshist $z$ are computed. 
This amounts to elimination of $z$ from them  to obtain the time profile of $a(t)$  
from the analysis of observational data. The obtained profile is shown in left
panel of Fig.\ \ref{fig:avst} and this corresponds to the 
redshift range $0 \leqslant z \leqslant 1$ or to the equivalent $t$-range: $0.44 \leqslant t(z) \leqslant 1$
as obtained from Eq.\ (\ref{eq:jla3}). $t$ is also normalised to unity  ($t_0 = 1$)
at present epoch $z=0$.

\section{$k-$essence scalar field model of dark energy with field-independent potential}
\label{seciia}
The field theoretic models of dark energy involve
suitably chosen Lagrangians contributing
to energy-momentum tensor in the Einstein's field equations.
In this paper, we  consider  dark energy described by 
a homogeneous scalar field $\phi$ whose dynamics is governed by a $k$-essence Lagrangian
with a field-independent (constant) potential.
Such models involve actions with non-canonical kinetic terms
and have phenomenologically rich properties.  
Theories with  
a non-canonical kinetic term was first proposed by Born and Infeld in
\cite{new1} to get rid of the infinite self-energy of the electron and 
were also investigated by Heisenberg in \cite{new2} 
in the context of physics of cosmic rays
and meson productions and further developed by Dirac in
\cite{new3}. Such non-canonical Lagrangians  have also been
investigated  in low energy effective string theory and in
D-branes models \cite{new4,new5,new6}. In  
cosmology, the kinetic energy driven acceleration was
first proposed in the context of cosmic inflation  \cite{ArmendarizPicon:1999rj} 
and also
applied to explain the late time cosmic acceleration \textit{i.e.}
dark energy, in  
 \cite{ArmendarizPicon:2000ah,ArmendarizPicon:2000dh,ArmendarizPicon:2005nz,Chiba:1999ka,ArkaniHamed:2003uy,Caldwell:1999ew}. 
%In this paper we have considered the dynamics of dark energy to
%be driven by  $k$-essence models. 
In this section, we present a 
brief outline of the general framework and features  of the $k-$essence model
that would be relevant in the context of this paper.
\\

The dynamics of the $k-$essence scalar field $\phi$,   which is 
minimally coupled to the gravitational field $g_{\mu\nu}$
is driven by the action
\begin{eqnarray}
S &=& \int d^4x \sqrt{-g}{\cal L}(X,\phi)
\label{keq:kaction}
\end{eqnarray}
where $X \equiv \frac{1}{2} g^{\mu\nu}\nabla_\mu\phi\nabla_\nu\phi$, 
$g$ is determinant of the metric
$g_{\mu\nu}$ and $\nabla_\mu$ denotes
covariant derivative. 
Variation of the action with respect to the field $g_{\mu\nu}$
gives the energy momentum tensor for the $k-$essence field as
\begin{eqnarray}
T_{\mu\nu} & \equiv & \frac{2}{\sqrt{-g}} \frac{\delta S}{\delta g^{\mu\nu}} = \frac{\partial{\cal L}}{\partial X} \nabla_\mu\phi \nabla_\nu\phi 
- g_{\mu\nu}{\cal L}
\label{keq:emtensor}
\end{eqnarray}
which when  compared 
with the form of that of perfect fluid (characterised by energy density
$\tilde{\rho}$ and pressure $\tilde{p}$):
$T_{\mu\nu} = (\tilde{\rho} + \tilde{p}) u_\mu u_\nu -\tilde{p} g_{\mu\nu}$   gives
\begin{eqnarray}
\tilde p = {\cal L}(\phi,X)\, \quad  , \quad \tilde\rho = 2X \frac{\partial \tilde p}{\partial X} - \tilde{p}
\label{keq:genp}
\end{eqnarray}
where $u_\mu$ is the effective  four-velocity  given by
$
u_\mu = {\rm sgn}(\partial_0 \phi) \frac{\nabla_\mu \phi}{2X}$. We take
the form of the non-canonical
Lagrangian in $k$-essence model   as 
\begin{eqnarray}
{\cal L}(\phi,X) =  V(\phi) F(X)
\label{keq:1}
\end{eqnarray}
Motivations for consideration of such restricted form of the
$k$-essence Lagrangian density also
comes from low energy effective string theory with
a suitable redefinition of the field discussed comprehensively in
\cite{new7,new8,new9}. For such $k-$essence Lagrangians in
Eq.\ (\ref{keq:1}), the pressure and energy density as
given in Eq.\ (\ref{keq:genp}) may be written as
\begin{eqnarray}
\tilde p =  V(\phi) F(X)\, \quad , \quad 
\tilde \rho = V(\phi) (2XF_X - F)\label{keq:13} 
\end{eqnarray}
where $F_X \equiv dF/dX$. In an expanding FRW universe, described by
the FRW scale factor $a(t)$, when the dark energy component  
does not interact with any other component of the universe,
the energy density and pressure due to this component satisfies
the continuity equation $\dot{\tilde\rho}  + 3H(\tilde\rho  + \tilde p ) = 0$,  which 
using with Eq.\ (\ref{keq:13}) one obtains
\begin{eqnarray}
\Big{[}\frac{d}{dt}V(\phi) \Big{]}(2XF_X - F) + 
V(\phi)\Big{[}\frac{d}{dt} (2XF_X - F) \Big{]} + 6V(\phi)HXF_X  &=& 0 
\label{keq:17}
\end{eqnarray}
Taking the $k$-essence field  in the FRW spacetime background to be 
homogeneous: $\phi(t,\vec{x}) \equiv \phi(t)$, we have
$X = (1/2)\dot{\phi}^2$, $\Big{[}\frac{d}{dt}V(\phi) \Big{]} = \Big{[}\frac{d}{d\phi}V(\phi) \Big{]} \dot{\phi}$ and 
$\Big{[}\frac{d}{dt} (2XF_X - F) \Big{]} = \Big{[}\frac{d}{dX} (2XF_X - F) \Big{]} \dot{\phi} \ddot{\phi}$.  From Eq.\ (\ref{keq:17}), we then have 
 the equation of motion for the $k-$essence field in the
evolving in FRW background spacetime as 
\begin{eqnarray}
(F_X + 2XF_{XX})  \ddot{\phi} + 3  H F_X  \dot{\phi}
  + (2XF_X - F)  \frac{V_\phi}{V}  &=& 0  
\label{keq:20}
\end{eqnarray}
where $F_{XX} = d^2F/dX^2$ and $V_\phi = dV/d\phi$.\\

A simple class of such $k-$essence models,
called  `purely'  kinetic $k$-essence models was investigated in
\cite{Scherrer:2004au}, where the Lagrangian
involves only the kinetic factor - a function of $X$, 
without having any explicit dependence on the field $\phi$.
This corresponds to $k-$essence models with Lagrangians given by
Eq.\  (\ref{keq:1}) with a constant potential term: $V(\phi)  \equiv V_0$, a constant. Such models have been shown to generate an
exponential cosmic inflation but remains unable to
explain
the graceful exit from the phase of inflation. However,
in the context of dark energy, these purely kinetic $k-$essence 
models have been shown to generate the observed  transition from
the phase of decelerated expansion to a phase of accelerated expansion,
during late time evolution of the universe.\\

For such models with constant $V(\phi)$, $V_\phi=0$ and using
$X = \dot{\phi}^2/2$, $\dot{X} = \dot{\phi}\ddot{\phi}$, 
the
equation of motion (\ref{keq:20}) 
for the $k-$essence field can be written as
\begin{eqnarray}
(F_X + 2XF_{XX})  \dot{X} + 6  HX F_X    &=& 0
\label{keq:22}
\end{eqnarray}
Writing $H = \dot{a}/a$ and changing  the independent variable from $t$
to the scale factor $a$, the above equation can be expressed  
in a differential form as
\begin{eqnarray}
 d \ln(XF_X^2) + d\ln a^6 &=& 0 
\label{keq:23}  
\end{eqnarray} 
which on integration gives a scaling relation: 
\begin{eqnarray}
XF_X^2 = Ca^{-6}
\label{keq:24}
\end{eqnarray}
where $C$ is the constant of integration. In the 
context of a field independent potential, such a form of scaling relation 
holds only when the dark energy component does not interact with
dark matter and continuity equation in each of these sectors are satisfied
without any source term. In the context of interacting DE-DM scenario, the Eq.\ (\ref{keq:24}) is no longer valid.
In the following sections, we obtained
 the modified scaling relation, in the interacting DE-DM scenario for 
 $k-$ essence model of Dark energy  with a
constant potential.

\section{Framework of interactive Dark energy - Dark matter model}
\label{seciii}
In this section we describe the theoretical framework of interacting DE-DM 
scenario in  FRW flat spacetime background filled with
an ideal fluid with its components as DE and DM.
We focus only on the late time era of cosmic evolution where we neglect the
contribution to energy density due to baryonic matter and radiation,
as supported by  small estimated value of their combined share $(\sim 4\%)$
in present universe based on measurements from  
WMAP \cite{Hinshaw:2008kr} and Planck \cite{Ade:2013zuv}
experiments. The conservation
of energy momentum tensor for the total dark fluid 
with interactions between its components   may be expressed through 
the non-conserving continuity equations,
\begin{eqnarray}
\Big{[} \dot{\rho}_{\rm dm} + 3H {\rho}_{\rm dm}\Big{]} &=& Q(t) 
\label{eq:contdm}\\
\Big{[} \dot{\rho}_{\rm de} + 3H ( \rho_{\rm de} +  p_{\rm de})\Big{]}
&=& -Q(t)
\label{eq:contde}
\end{eqnarray}
where, $\rho_{\rm de}$ and $\rho_{\rm dm}$ denote the instantaneous
energy densities of DE and DM components respectively and
$p_{\rm de}$ is the pressure of  DE fluid. Dark matter is considered
to be (non-relativistic) pressure-less  dust.
The function $Q(t)$ in the source term of the above equation
gives a measure of instantaneous rate of
energy transfer between DE and DM components.
Eqs.\ (\ref{eq:contdm}) and (\ref{eq:contde}) imply
\begin{eqnarray}
\dot{\rho}  + 3H(\rho + p) & = &0
\label{eq:conttot}
\end{eqnarray}
which implies  conservation of energy momentum tensor of the total dark fluid with
energy density $\rho = \rho_{\rm de} + \rho_{\rm dm}$ and
pressure $p = p_{\rm de}$. The dot ($\cdot$) represents 
derivatives with respect to the dimensionless time parameter
$t$ which is normalised to $t=1$ at present epoch. As discussed
in Sec.\ \ref{secii}, the time profile of $a(t)$ has been
extracted from analysis of observational data over the
late time domain $0.44 \lsim t \leqslant 1$  accessible in SNe Ia observations.
We express the time dependence of the source term  $Q(t)$ through the scale 
factor $a(t)$ and a constant parameter $k$ as
\begin{eqnarray}
Q(t)=Q_0  \big{[}a(t) \big{]} ^k
\label{eq:qt}
\end{eqnarray}
Where $Q_0$ is the value of $Q(t)$ at present epoch since we
used the normalisation $a(t)=1$  at present epoch. 
For convenience, we express time dependences of various quantities
in terms of a  time parameter $\eta$ defined as
$\eta \equiv \ln a(t)$ ($\eta=0$ then corresponds to the present epoch and $a = e^{\eta}$). 
Eqs.\ (\ref{eq:contdm}) and  (\ref{eq:contde}) then take the form
\begin{eqnarray}
\rho_{\rm dm}^\prime + 3\rho_{\rm dm} &=& Q_0\frac{e^{k\eta}}{H}
\label{eq:contetadm}\\
\rho_{\rm de}^\prime + 3(\rho_{\rm de} + p_{\rm de}) &=& 
- Q_0\frac{e^{k\eta}}{H}
\label{eq:contetade}
\end{eqnarray}
where all the time-dependent quantities involved, 
are regarded as functions of $\eta$ and 
$^\prime$ denotes derivative with respect to $\eta$.
Multiplying both sides of Eq.\ (\ref{eq:contetadm}) by $e^{3\eta}$
and writing the right hand side as a total derivative we get
\begin{eqnarray}
\dv{}{\eta}  \Big{[} e^{3\eta}\rho_{\rm dm} \Big{]}
&=& Q_0\frac{e^{(k+3)\eta}}{H}
\label{eq:totder}
\end{eqnarray}
which on integration between limits $\eta=0$ and $\eta=\eta$ gives
\begin{eqnarray}
\rho_{\rm dm}(\eta)  
&=&
e^{-3\eta}\left[\rho_{\rm dm}^0 + Q_0\int_0^\eta \frac{d\eta_1\ 
e^{(k+3)\eta_1} }{H(\eta_1)}\right]\,
\label{eq:rhodmeta}
\end{eqnarray}
where,  $\rho_{\rm dm}^0$ denotes the DM density at
present epoch. Note that, in absence of any interaction
between DE and DM ($Q_0=0$),  we
get $\rho_{\rm dm}(\eta) = e^{-3\eta}\rho_{\rm dm}^0 
= a^{-3}\rho_{\rm dm}^0$, which is as expected from $\Lambda$-CDM model.
Dividing both sides by the total dark fluid density at present epoch, $(\rho_{\rm de}^0 + \rho_{\rm dm}^0)$,
we have
\begin{eqnarray}
\frac{\rho_{\rm dm}(\eta;k,\beta_0)}{(\rho_{dm}^0 + \rho_{de}^0)}
&=& 
e^{-3\eta}\left[\Omega_{\rm dm}^0  
+ 
\beta_0 
 \int_0^\eta \frac{d\eta_1\ 
e^{(k+3)\eta_1} }{H(\eta_1)}
\right] 
\label{eq:rhodmeta1a}
\end{eqnarray}
Here, $\beta_0 \equiv Q_0 /(\rho_{\rm dm}^0 + \rho_{\rm de}^0)$ and 
$\Omega_{\rm dm}^0 (\equiv \rho_{\rm dm}^0/(\rho_{\rm dm}^0 + \rho_{\rm de}^0))$ 
denotes fractional content of DM in the present universe whose value
is $\sim 0.268$ from WMAP and PLANCK observations \cite{Hinshaw:2008kr, Ade:2013zuv}.
$k$ and $\beta_0$ are also put in the argument of  $\rho_{\rm dm}$ to emphasize the 
point that 
temporal behaviour of the DM energy density, in this scenario,  
depends on values of these parameters. \\

The time profile of the Hubble parameter $H(t)=\dot{a}/a$ may be obtained 
from the  temporal profile of $a$ and $\dot{a}$ as shown in Fig.\ \ref{fig:avst}.
Numerically eliminating $t$ from the $a(t)$ and $H(t)$
we may express the Hubble parameter as a function
of $a$ or subsequently as a function of $\eta$ as $H(\eta)$  
which appear in the right hand side of Eq.\ (\ref{eq:rhodmeta1a}).
We find that the quantity $1/H(\eta)$, thus obtained, corresponding
to the central line  in Fig. \ref{fig:EZ}, may be fitted with a polynomial of order
3, which we express as
\begin{eqnarray}
\frac{1}{H(\eta)} = \sum_{m=0}^3 A_m \eta^m \quad  
\mbox{where, }  A_0 = 1, A_1 = 0.45, A_2 = -0.32, A_3 = -0.15.
\label{eq:onebyh}
\end{eqnarray}
Using Eq.\ (\ref{eq:onebyh}) in the integral appearing in 
right hand side of  Eq.\ (\ref{eq:rhodmeta1a}) we have
\begin{eqnarray}
 \frac{\rho_{\rm dm}(\eta;k,\beta_0)}{(\rho_{dm}^0 + \rho_{de}^0)}  &=&
e^{-3\eta}\left[\Omega_{\rm dm}^0 + \beta_0\sum_{m=0}^3A_m  I_{k,m}(\eta) \right]
\label{eq:rhodmeta1}
\end{eqnarray}
where  
\begin{eqnarray}
I_{k,m}(\eta) & \equiv & \int_0^\eta d\eta_1\ e^{(k+3)\eta_1}{(\eta_1)}^m
\label{eq:Ifunc}
\end{eqnarray}
For $k \neq -3$, the function $I_{k,m}(\eta)$ satisfies the recursion relation
\begin{eqnarray}
I_{k,m}(\eta)  
&=&    \frac{\eta^m e^{(k+3)\eta}}{k+3}  
- \frac{m}{k+3} I_{k,m-1}(\eta)\label{eq:Ifuncrecur}   \\
&&\mbox{with} \quad I_{k,0}(\eta) = \int_0^\eta d\eta_1 e^{(k+3)\eta_1} 
= \frac{e^{(k+3)\eta}}{k+3} \nonumber
\label{eq:Ifunc0}
\end{eqnarray}
However, for $k = -3$, the exponential term in Eq.\ (\ref{eq:Ifunc}) becomes unity
and we can easily compute the integral as
\begin{eqnarray}
I_{-3,m}(\eta) & \equiv & \int_0^\eta d\eta_1\ {(\eta_1)}^m
= \frac{\eta^{m+1}}{m+1}\,
\label{eq:Ifunckm3}
\end{eqnarray}
where $m$ can take 4 values \textit{viz.} 0, 1, 2, 3 as evident from
Eq.\ (\ref{eq:onebyh}).
Using Eq.\ (\ref{eq:Ifuncrecur}) or (\ref{eq:Ifunckm3}), according as $k \neq -3$ or $k = -3$,
we can write the term $\displaystyle\sum_{m=0}^3A_m  I_{k,m}(\eta)$ appearing in right hand side
of Eq.\ (\ref{eq:rhodmeta1}) as 
\begin{eqnarray}
\sum_{m=0}^3A_m  I_{k,m}(\eta)
&=&
\begin{cases}
(B_{k0} + B_{k1} \eta + B_{k2} \eta^2 +  B_{k3} \eta^3 )e^{(k+3)\eta} \quad \mbox{for } k \neq -3\\
A_0\eta + \frac{A_1\eta^2}{2}  + \frac{A_2\eta^3}{3} +  \frac{A_3\eta^4}{4} \quad \mbox{for } k = -3
\end{cases}
\label{eq:sumterm}
\end{eqnarray}
where, the constants $B_{ki}$'s can be expressed as
\begin{eqnarray*}
B_{k0} &=& \left(\frac{A_0}{k+3} - \frac{A_1}{(k+3)^2} + \frac{2A_2}{(k+3)^3} -   \frac{6A_3}{(k+3)^4}\right)\\
B_{k1} &=&\left(\frac{A_1}{k+3} - \frac{2A_2}{(k+3)^2} +  \frac{6A_3}{(k+3)^3} \right)\\
B_{k2} &=& \left( \frac{A_2}{k+3} - \frac{3A_3}{(k+3)^2} \right)\\
B_{k3} &=& \left(\frac{A_3}{k+3}\right)
\end{eqnarray*}
With the aid of Eq.\ (\ref{eq:sumterm}) the $\eta$-dependence appearing
in right hand side of Eq.\ (\ref{eq:rhodmeta1}) 
 can be expressed in an algebraic form  in terms of parameters $k$ and
$\beta_0$,  with known values of all other factors involved in the expression,  
\textit{e.g} $A_i$'s and $\Omega_{\rm dm}^0 \sim 0.268$. This allows us
to numerically compute the time($\eta$)-profile of DM energy density term
for any chosen benchmark values of the parameters: $k$ and $\beta_0$.
Note that, determination of   values of  constants $A_i$'s (and so also $B_{ki}$'s for any given $k$)
uses time profile of   scale factor as extracted from analysis of SNe Ia data. 
The obtained $\eta$-dependence of  DM energy density term 
 in   DE-DM interacting scenario  is, therefore, consistent with
the Sne Ia data. The features of the data
 are encoded in the corresponding expression through the 
constants $A_i$'s. We may now also use
the substitution $\eta = \ln a$ in Eqs.\ (\ref{eq:rhodmeta1}) and (\ref{eq:sumterm}) 
to express the time dependence in terms of scale factor $a$ itself as
\begin{eqnarray}
 \frac{\rho_{\rm dm}(a;k,\beta_0)}{(\rho_{dm}^0 + \rho_{de}^0)}  &=&
a^{-3}\left[\Omega_{\rm dm}^0 + \beta_0\sum_{m=0}^3A_m  I_{k,m}(\ln a) \right]
\label{eq:rhodmvsa}
\end{eqnarray}

The total energy density $\rho  = \rho_{\rm dm}+\rho_{\rm de}$
and pressure $p = p_{\rm de}$
of the dark fluid, on the other hand, are
independent of the parameters $\beta_0$ and $k$, as  
the continuity equation of the total dark fluid involves no source term.
 We can obtain their temporal behaviour,
directly from the time profile of the scale factor obtained from the analysis
discussed in Sec.\ \ref{secii}. To see this, we write the Friedmann equations
governing late-time cosmic evolution, which in a
flat FRW spacetime background with DE and DM as its primary contents
take the forms 
\begin{eqnarray}
H^2 
&=& \left(\frac{\dot{a}}{a}\right)^2 = \frac{\kappa^2}{3} (\rho_{\rm dm} + \rho_{\rm de})
\label{eq:fri1}\\
\frac{\ddot{a}}{a} 
&=& -\frac{\kappa^2}{6} \Big{[} (\rho_{\rm dm} + \rho_{\rm de}) + 3p_{\rm de}\Big{]}
\label{eq:fri2}
\end{eqnarray}
where $\kappa^2 \equiv 8\pi G$ ($G$ is the Newton’s Gravitational constant).
Using Eqs.\ (\ref{eq:fri1}) and  (\ref{eq:fri2}), we may express the 
 equation of state $w$ of the total dark fluid   in terms of the scale factor 
and its   time derivatives as
\begin{eqnarray}
w & \equiv & \frac{p_{\rm de}}{\rho_{\rm de} + \rho_{\rm dm}}
= -\frac{2}{3}\frac{a\ddot{a}}{\dot{a}^2} - \frac{1}{3}
\label{eq:eostotal}
\end{eqnarray}
From the obtained time profile of scale factor $a(t)$ and its derivatives as shown in 
Fig.\ \ref{fig:avst} we can obtain $t-$dependence of $w(t)$.  
Numerically eliminating $t$ from the $a(t)$ and $w(t)$
we may express the  equation of state parameter $w$ as a function
of $a$  or  $\eta$, making use of the substitution $\eta = \ln a$.
We find that, $w(\eta)$ thus obtained, corresponding
to the central line in  Fig.\ \ref{fig:EZ}, may be fitted with a polynomial of order 5.
This we express as 
\begin{eqnarray}
w(\eta) &=& -1 + \sum_{i=0}W_i \eta^i \label{eq:wetafit}
\end{eqnarray}
with values of the coefficients $W_i$ at best-fit are given by
\begin{eqnarray}
&& W_0 = -0.70\, , W_1 = -0.61\,, W_2 = -0.49\,,
W_3 = -2.29\,,  \nonumber\\
&& W_4 = -2.81\,, W_5 = -0.92\,,  
\mbox{ and }W_i = 0\, \mbox{ for } i>5
\label{eq:wetacoeff}
\end{eqnarray}

In terms of the parameter $\eta$ the 
continuity equation (\ref{eq:conttot}) for the total dark fluid takes the form
\begin{eqnarray}
\frac{d}{d\eta} \ln \Big{(} \rho_{\rm dm} + \rho_{\rm de}\Big{)}
&=& - 3 \Big{(} 1 + w(\eta)\Big{)}
\label{eq:conttoteta}
\end{eqnarray}
which on integration between the limits $\eta=0$ and $\eta=\eta$ gives
\begin{eqnarray}
\frac{(\rho_{\rm de}  + \rho_{\rm dm} )_{\eta}}{(\rho_{\rm de}^0 + \rho_{\rm dm})_0}
&=&
\exp\left[-3\int_{0}^\eta \big{(}1 + w(\eta_1)\big{)}d\eta_1 \right] 
\label{eq:rhoetacoeff}
\end{eqnarray}
The integral within  exponent
can be performed by using Eq.\ (\ref{eq:wetafit}), with  $W_i$'s  as given  
in Eq.\ (\ref{eq:wetacoeff}). This gives
\begin{eqnarray}
\frac{ \rho_{\rm de}(\eta) + \rho_{\rm dm} (\eta) }{(\rho_{\rm de}^0 + \rho_{\rm dm}^0)}
&=&
\exp\left[-3\int_{0}^\eta \left( \sum_{i=0}^5 W_i \eta_1^i  \right)d\eta_1 \right]  
%&=&
%\exp\left[-3\left( \sum_{i=0}^5 W_i \int_{0}^\eta \eta_1^i d\eta_1 \right)  \right] \nonumber\\
= 
\exp\left[-3\left( \sum_{i=0}^5  \frac{W_i   \eta^{i+1}}{i+1}  \right)  \right]  
\label{eq:rhototeta1}
\end{eqnarray}
We may again use the substitution $\eta = \ln a$ in the above
to express the  energy density of the total dark fluid $\rho = \rho_{\rm de} + \rho_{\rm dm}$
as a function of scale factor $a$ as,
\begin{eqnarray}
\frac{\rho(a)}{\rho_{\rm de}^0 + \rho_{\rm dm}^0}
&=&
\exp\left[-3\left( \sum_{i=0}^5  \frac{W_i   (\ln a)^{i+1}}{i+1}  \right)  \right]  
\label{eq:rhototvsa}
\end{eqnarray}
We find that this scale factor dependence of the total energy density,
thus obtained,  can be fitted best with a fourth order polynomial expressed  as,
\begin{eqnarray}
\frac{\rho(a)}{\rho_{\rm de}^0 + \rho_{\rm dm}^0} 
&=&   \sum_{m=0}^4 R_m a^m   
\label{eq:rhotot}
\end{eqnarray}
with the best-fit values of coefficients as 
\begin{eqnarray}
R_0 = 29.3, \quad R_1 = -120.3, \quad R_2 = 199.96, 
\quad R_3 = -151.75, \quad R_4 = 43.8
\label{eq:Rval}
\end{eqnarray}
 Note that,  the constants $W_i$'s and $R_i$'s
  encode the inputs from the observational data
used here.    \\

\begin{figure}[t]	
\centering
\includegraphics[width=0.7\linewidth]{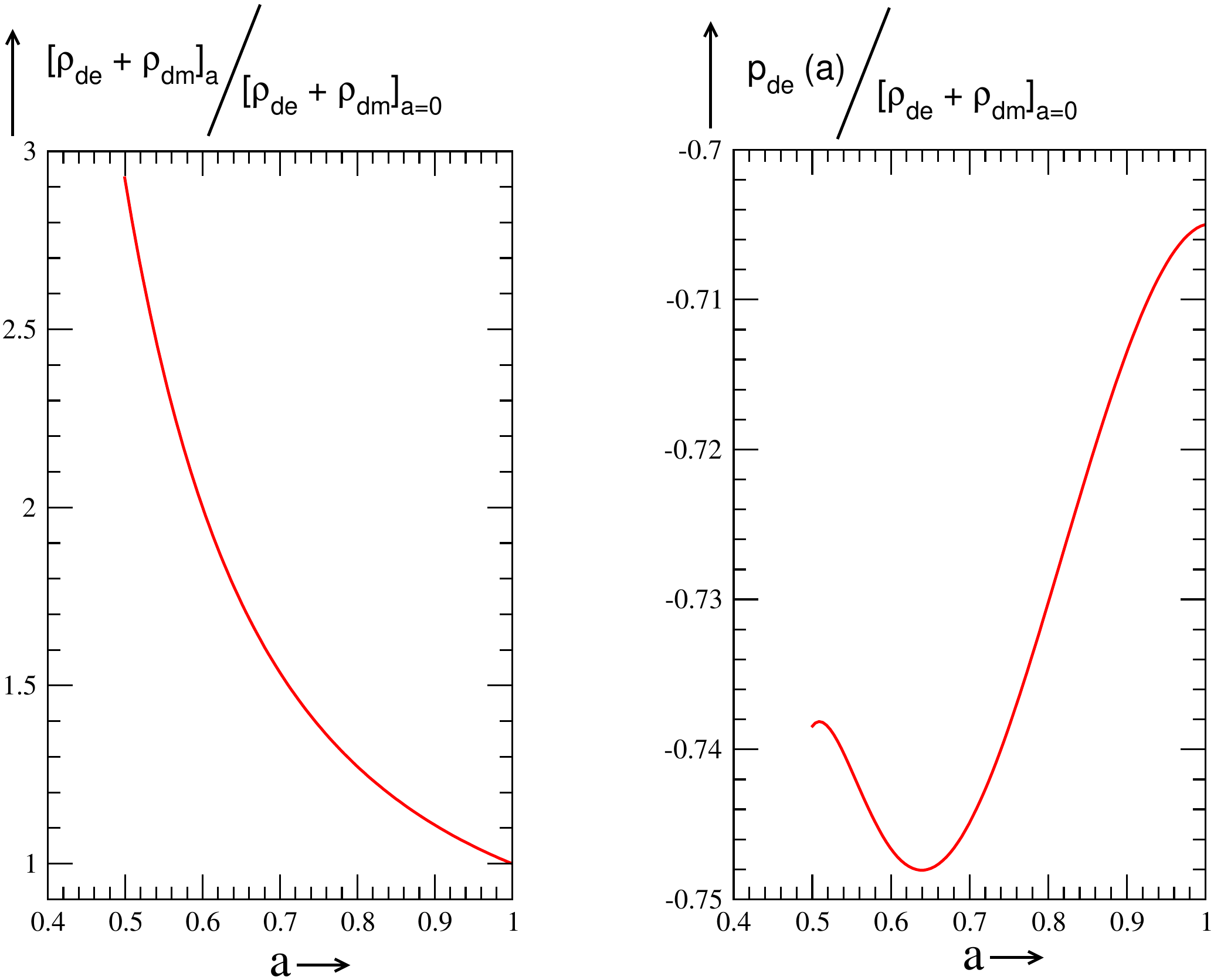} 
\caption{Temporal behaviour of cosmological parameters obtained from SNe Ia observation (Left panel: Total energy density as a function of scale factor, Right panel: Pressure of dark sector as a function of scale factor). }
\label{fig:obs}
\end{figure}

Similarly we may also obtain temporal behaviour of the pressure  ($p_{\rm de}$) of 
DE. Since dark matter dust has zero
pressure, the equation of state parameter of the dark fluid can be written as
\begin{eqnarray}
w(\eta) &=& \frac{p_{\rm de}(\eta)}{\rho_{\rm de}(\eta) + \rho_{\rm dm}(\eta)}\,,
\end{eqnarray}
from which we may write 
\begin{eqnarray}
\frac{p_{\rm de}(\eta)}{ \rho_{\rm de}^0 + \rho_{\rm dm}^0}
&=& 
w(\eta) \frac{\rho_{\rm de}(\eta) + \rho_{\rm dm}(\eta)}{ \rho_{\rm de}^0 + \rho_{\rm dm}^0}
\label{eq:pdeform1}
\end{eqnarray}
Using Eqs.\ (\ref{eq:wetafit}) and (\ref{eq:rhototeta1})
in Eq.\ (\ref{eq:pdeform1}) and by making the substitution $\eta = \ln a$
we may express this functional behaviour of the $p_{\rm de}$
in terms of $a$  as 
\begin{eqnarray}
\frac{p_{\rm de}(a)}{ \rho_{\rm de}^0 + \rho_{\rm dm}^0}
&=& 
\Big{[} -1 + \sum_{i=0}^5 W_i (\ln a)^i \Big{]} \cdot
\exp\left[-3\left( \sum_{i=0}^5  \frac{W_i   (\ln a)^{i+1}}{i+1}  \right)  \right] 
\label{eq:pdeform2}
\end{eqnarray}
We find that, this scale factor dependence of $p_{\rm de}$  
can be fitted best with a  polynomial of order 6 expressed  as,
\begin{eqnarray}
\frac{p_{\rm de}(a)}{\rho_{\rm de}^0 + \rho_{\rm dm}^0} 
&=&   \sum_{m=0}^6 P_m a^m   
\label{eq:pdeafit}
\end{eqnarray}
with the best-fit values of coefficients as 
\begin{eqnarray}
&& \quad\quad\quad\quad P_0 = -11.84 , \quad P_1 = 90.04,  
\quad P_2 =  -297.92, \nonumber\\
&& P_3 = 515.97, \quad P_4 =  -495.22 , \quad P_5 = 250.83,  \quad P_6 = -52.57
\label{eq:Pval}
\end{eqnarray}
The   scale factor dependence of   $\rho(a)$ 
and   $p_{\rm de}(a)$, thus obtained
using the temporal profile of $a(t)$   from SNe Ia data, are 
shown in left panel and right panel of Fig.\ \ref{fig:obs} respectively.\\

\begin{figure}[t]	
\centering
\includegraphics[scale=0.3]{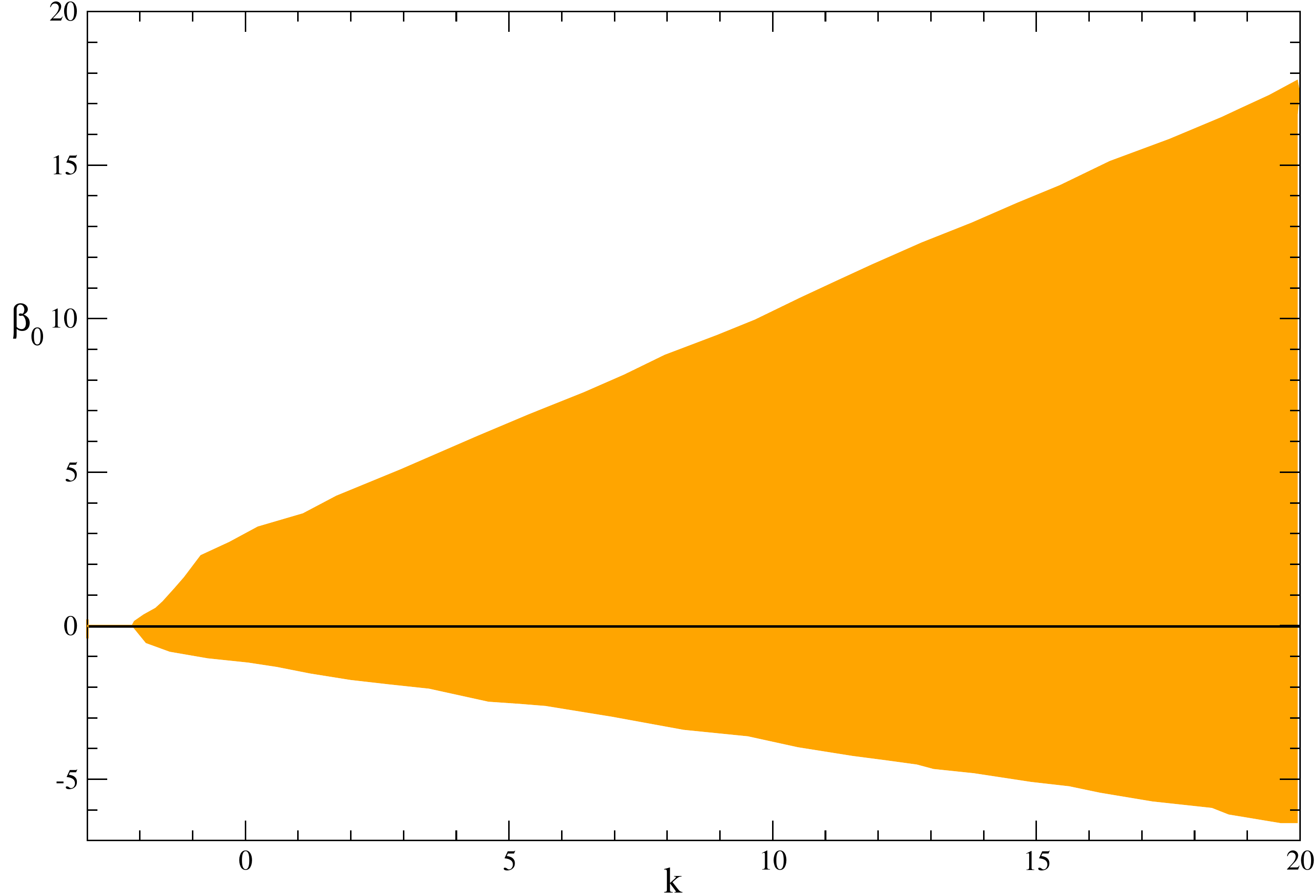} 
\caption{Range of parameter space of model parameters $k$ vs. $\beta_0$
for which the condition in Eq.\ (\ref{eq:kbeta2}) is satisfied. A black line is drawn at $\beta_0=0$ to reflect the fact that for $\beta_0 = 0$, the parameter $k$ loses its relevance in the context (described in text).}
\label{fig:param}
\end{figure}

Note that, the density of DE  as a function of the scale factor $a$ may now 
be expressed as 
\begin{eqnarray}
\frac{\rho_{\rm de}(a;k,\beta_0)}{(\rho_{dm}^0 + \rho_{de}^0)}
&=& 
\frac{\rho(a)}{(\rho_{dm}^0 + \rho_{de}^0)}
- \frac{\rho_{\rm dm}(a;k,\beta_0)}{(\rho_{dm}^0 + \rho_{de}^0)}\,,
\label{eq:rhodeeta3}
\end{eqnarray}
which   involves the parameters $\beta_0$ and $k$, due 
to their appearance in $\rho_{\rm dm}(a;k,\beta_0)$. Since
energy density is always a positive quantity and 
the scale factor or time dependence 
of  energy density of total dark fluid has
already been obtained directly from observation
(left panel of Fig.\ \ref{fig:obs}),
the estimated profile of the dark matter density 
$\rho_{\rm dm}(a,k,\beta_0)$ computed from 
Eq.\ (\ref{eq:rhodmeta1a}) for given values of $k$ and
$\beta_0$ are subject to the constraint
\begin{eqnarray}
 0 < \frac{\rho_{\rm dm}(a;k,\beta_0)}{(\rho_{\rm de}^0 + \rho_{\rm dm}^0)} 
 < \frac{\rho(a)}{(\rho_{\rm de}^0 + \rho_{\rm dm}^0)}
\label{eq:kbeta1}
\end{eqnarray}
for the accessible domain of $a$ ($0.5 \lesssim a < 1$) in SNe Ia observations. Using 
Eqs.\ (\ref{eq:rhodmvsa}) and (\ref{eq:rhotot}), we may express the above condition
as 
\begin{eqnarray}
0 < a^{-3}\left[\Omega_{\rm dm}^0 + \beta_0\sum_{m=0}^3 A_m  I_{k,m}(\ln a) \right]  <
\displaystyle\sum_{m=0}^4 R_m a^m
\label{eq:kbeta2}
\end{eqnarray}
where the function $I_{k,m}(\ln a)$ contains the parameter $k$ as
seen from its explicit form in Eq.\ (\ref{eq:sumterm})  with $\eta = \ln a$.
Using the form of  $I_{k,m}(\ln a)$ for $k \neq -3$,  
we find the range in the parameter space spanned by $k$ and $\beta_0$,
every point ($k,\beta_0$) of which satisfies the condition in
Eq.\ (\ref{eq:kbeta2}). This allowed region 
in   parameter space has been shown by the shaded 
region in Fig.\ \ref{fig:param}. However, for $k=-3$
(when $Q(t) = Q_0 a^{-3})$, we find the range of $\beta_0$
 for which the condition in Eq.\ (\ref{eq:kbeta2})
is satisfied, is  $-0.2 \leqslant \beta_0 \leqslant 0.42$. \\ 

\begin{figure}[t]	
\centering
\includegraphics[scale=0.5]{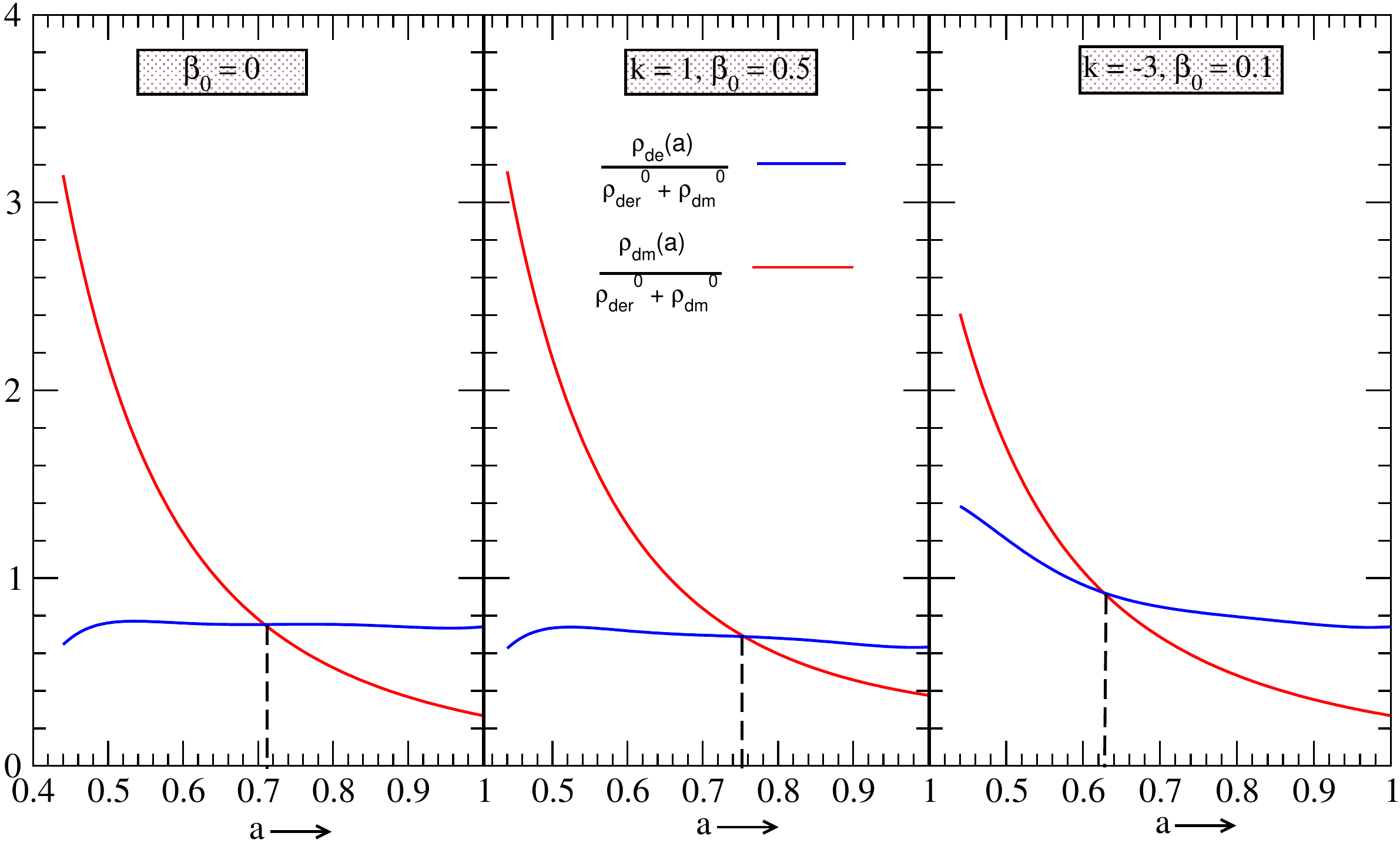} 
\caption{Variation of energy density of dark  energy and dark matter
with scale factor for chosen benchmark values of the parameters ($k$ and $\beta_0$).}
\label{fig:ET}
\end{figure}
For some benchmark values of $k$ and $\beta_0$
chosen within this allowed region we have also
depicted the variation of the dark energy and dark matter densities 
with the scale factor in Fig.\ \ref{fig:ET}. Note from Eq.\ (\ref{eq:qt}) that for $\beta_0 = 0 = Q_0$, the parameter $k$ becomes irrelevant
and corresponding continuity equation for DM (Eq.\ (\ref{eq:contdm})) has no source
term ($Q(t)=0$). This implies $\rho_{\rm dm}(a) = \rho_{\rm dm}^0 a^{-3}$ with  
$\rho_{\rm de}(a) = \rho(a) - \rho_{\rm dm}^0 a^{-3}$, where
$\rho(a)$ as directly obtained from observation
has been expressed
thorough a fitted polynomial in Eq.\ (\ref{eq:rhotot}). These  profiles   
for $\beta_0=0$ (which corresponds to `non interacting DE-DM' scenario) are presented
in left panel  of Fig.\ \ref{fig:ET}, 
and we find that
DM energy density falling as $\sim 1/a^3$ falls below dark energy
density at an epoch corresponding to $a \sim 0.71$.
For $\beta_0=0.5$ and $k=1$ (middle panel of Fig.\ \ref{fig:ET})
the   source term $Q(t)$ grows with time having the same profile as that $a(t)$ itself,
and the corresponding DM and DE density profiles 
 evaluated using Eqs.\ (\ref{eq:rhodmvsa}) and 
(\ref{eq:rhodeeta3}) show  that  DM energy density  falls below
that of DE at an epoch when $a \sim 0.75$.
For $\beta_0=0.1$ and $k=-3$  
the source term $Q(t)$ falls as $\sim 1/a(t)^3$ as  the universe expands,
and the corresponding plots presented in right panel of Fig.\ \ref{fig:ET} 
show beginning of dominance of DE density
over that of DM at a relatively earlier epoch marked by $a \sim 0.63$.

\section{$k$-essence model with constant potential and modification of scaling relation in presence of DE-DM interaction}
\label{seciv}
We assume dynamics of  dark energy of  universe to
be driven by a homogeneous scalar field $\phi = \phi(t)$ 
governed by  non-canonical $k-$essence Lagrangian 
 of the form $L = V(\phi)F(X)$ where the potential $V(\phi)=V_0$
is considered to be constant   and the dynamical term $F(X)$ is a function of $X 
\equiv (1/2)g_{\mu\nu}\nabla^\mu\phi\nabla^\nu\phi = (1/2)\dot{\phi}^2$.
The  energy density and pressure of DE are identified with the corresponding
quantities in the context of the $k$-essence model 
which may be expressed as  
\begin{eqnarray}
\rho_{\rm de} &=& V_0 (2XF_X - F)  \label{eq:krhode}\\
p_{\rm de} &=& V_0 F(X)  \label{eq:kpde}
\end{eqnarray}
where, $F_X \equiv dF/dX$. In the context of 
this constant potential $k-$essence model, it has been shown in \cite{Scherrer:2004au,dePutter:2007ny} that,
in the non-interacting DE-DM scenario when the
energy conservation is separately conserved in the DE sector (and also in DM sector) implying  
corresponding continuity equation (\ref{eq:contde}) being satisfied
 with $Q(t) = 0$,
we have a scaling relation of the form $XF_X^2 = Ca^{-6}$  where $C$ is a constant. 
In this section we show that, even in presence of time-dependent interactions
implying continual exchange of energy between the two sectors ($Q(t) \neq 0$), 
the constancy of   $k$-essence potential may lead to a modified
form of the scaling. Below we discuss how we obtain  
the modified form of scaling taking into 
consideration  observational inputs from SNe Ia data.\\

To obtain this we use Eqs.\ (\ref{eq:krhode}) and (\ref{eq:kpde}) in 
Eq.\ (\ref{eq:contde}) with $Q(t)$ parametrised as $Q(t) = Q_0 [a(t)]^k$ (Eq.\ (\ref{eq:qt})) to get
\begin{eqnarray}
\dv{}{t}  \Big{[} V_0 (2XF_X - F) \Big{]} 
+ 3H (2V_0XF_X) &=& -Q_0 a^k\,.
\end{eqnarray}
Writing $\displaystyle\frac{d}{dt} = \dot{a}\frac{d}{da}$ and $H = \dot{a}/a$,  after some rearrangements 
the above equation takes the form
\begin{eqnarray}
%\frac{(2XF_{XX}+F_{X})}{XF_{X}}dX + 6 \frac{da}{a} &=& -\frac{Q_0}{V_0} \frac{a^{k-1}}{H}\frac{da}{XF_{X}} \nonumber\\
 d \Big{[}\ln\{a^6(XF_X^2)\}\Big{]}  &=& - \frac{Q_0}{V_0} \frac{a^{k-1}}{H}\frac{da}{(XF_X)}
\label{eq:sc1}
\end{eqnarray}
where $F_{XX}=d^2F/dX^2$ and the quantities  $H$, $X$, $F$ and its derivatives 
are regarded as functions of scale factor $a$. Integrating
both sides of Eq.\ (\ref{eq:sc1}) between the limits $a=1$ (present epoch) 
and $a=a$ we have,
\begin{eqnarray}
&& \Big{[}\ln\{a^6(XF_X^2)\}\Big{]} -  \Big{[}\ln\{(XF_X^2)\}\Big{]}_{a=1}
= 
- \frac{Q_0}{V_0}\int_1^a\frac{ f(a')da'}{(XF_X)}  \nonumber\\
\mbox{or,} && \quad\quad 
XF_X^2 = C a^{-6} \exp\left(- \frac{Q_0}{V_0}\int_1^a\frac{ f(a';k)da'}{(XF_X)} \right)
\label{eq:sc2}
\end{eqnarray}
where, $C \equiv XF_X^2\big{|}_{a=1}$ is a constant and 
$f(a;k) \equiv \frac{a^{k-1}}{H(a)}$. In Eq.\ (\ref{eq:onebyh})
we have shown how $H^{-1}$ depends on $\eta = \ln a$, with the help of
a polynomial of $\ln a$, with given of values of coefficients $A_i$'s
encoding the observational inputs extracted from the analysis of JLA data.
We may, equivalently, fit this scale factor dependence of  $H^{-1}$ 
by a polynomial in $a$ (instead of $\ln a$) and found that, the
corresponding
best-fit curve may be given by a second order
polynomial of scale factor $a$ as
\begin{eqnarray}
\frac{1}{H(a)} &=& \sum_{n=0}^2 D_n a^n
\quad
\mbox{with } D_0 = -0.272\,, D_1 = 2.174\,, D_2 = -0.9.
\label{eq:onebyhvsa}
\end{eqnarray}
To perform the integration involved in Eq.\ (\ref{eq:sc2}), we can then
express the function $f(a,k)$ as a polynomial
\begin{eqnarray*}
f(a;k) &=& a^{k-1} (D_0 + D_1 a + D_2 a^2)\,,
\label{eq:fa}
\end{eqnarray*}
and we also need to express $(XF_X)$ as function of $a$.
To do so, we eliminate $F(X)$ from Eqs.\ (\ref{eq:krhode}) and   (\ref{eq:kpde})
and write
\begin{eqnarray}
XF_X  &=&  \frac{1}{2V_0} \Big{[}\rho_{\rm de} + p_{\rm de}  \Big{]}
\label{eq:xfx1}
\end{eqnarray}
Using Eq.\ (\ref{eq:rhodeeta3}) and writing all variables and parameters explicitly, we may write the above equation as
\begin{eqnarray}
XF_X  &=&  \frac{(\rho_{\rm dm}^0 + \rho_{\rm de}^0)}{2V_0} 
\Bigg{[}  \frac{\rho(a)}{(\rho_{\rm dm}^0 + \rho_{\rm de}^0)}
- \frac{\rho_{\rm dm}(a;k,\beta_0)}{(\rho_{dm}^0 + \rho_{\rm de}^0)}
+ \frac{p_{\rm de}(a)}{(\rho_{\rm dm}^0 + \rho_{\rm de}^0)}  \Bigg{]} \nonumber\\
& = &
 \frac{(\rho_{\rm dm}^0 + \rho_{\rm de}^0)}{2V_0} g(a;k,\beta_0)
\label{eq:xfx2}
\end{eqnarray}
where
\begin{eqnarray}
g(a; k,\beta_0) 
& \equiv &
\Bigg{[}  \frac{\rho(a)}{(\rho_{\rm dm}^0 + \rho_{\rm de}^0)}
- \frac{\rho_{\rm dm}(a;k,\beta_0)}{(\rho_{dm}^0 + \rho_{\rm de}^0)}
+ \frac{p_{\rm de}(a)}{(\rho_{\rm dm}^0 + \rho_{\rm de}^0)}  \Bigg{]} 
\label{eq:g}
\end{eqnarray}
Finally, using Eq.\ (\ref{eq:xfx2}) in Eq.\ (\ref{eq:sc2}) we obtain
\begin{eqnarray}
\frac{XF_X^2}{Ca^{-6}} &=& \exp\left(-2\beta_0\int_1^a\frac{ f(a_1;k)da_1}{g(a_1; k,\beta_0) } \right)
\label{eq:sc3}
\end{eqnarray}
All the three terms in the right hand side of Eq.\ (\ref{eq:g})
have been expressed in algebraic form in Eqs.\ (\ref{eq:rhodmvsa}),
(\ref{eq:rhotot}) and (\ref{eq:pdeafit}), using which
we can numerically compute the function  $g(a; k,\beta_0)$ for any
input values of $a, k, \beta_0$. 
With this and using the form of the function $f(a;k)$ in Eq.\ (\ref{eq:fa}),
we may numerically evaluate the integral within the exponent appearing
in Eq.\ (\ref{eq:sc3}).
This is the modified scaling relation arising 
out of the constancy of the potential $V=V_0$ of  $k$-essence model of dark 
energy, in presence of interaction between DE and DM
parametrised in terms of $\beta_0$ and $k$. 
The inputs from   observational data are encoded in the form
of the functions $f(a;k)$ and $g(a; k,\beta_0)$ 
through the various coefficients $D_i$'s, $A_i$'s, $P_i$'s, $R_i$'s \textit{etc.}
introduced  in Secs.\ \ref{seciii} and \ref{seciv},
while establishing connections  
with the temporal profile of the scale factor obtained from the
SNe Ia data in Sec.\ \ref{secii}. In the context of this constant
potential $k-$essence model of dark energy in interacting DE-DM scenario, the modified scaling
relation (\ref{eq:sc3}) establishes a connection between
the dynamical terms $X$, $F(X)$ involved in the $k-$essence Lagrangian
and the scale factor $a(t)$ of FRW universe along with the  parameters $k$ and $\beta_0$.
The constancy of the $k$-essence potential
is instrumental in establishing the relation.\\

\begin{figure}[t]	
\centering
\includegraphics[width=0.8\linewidth]{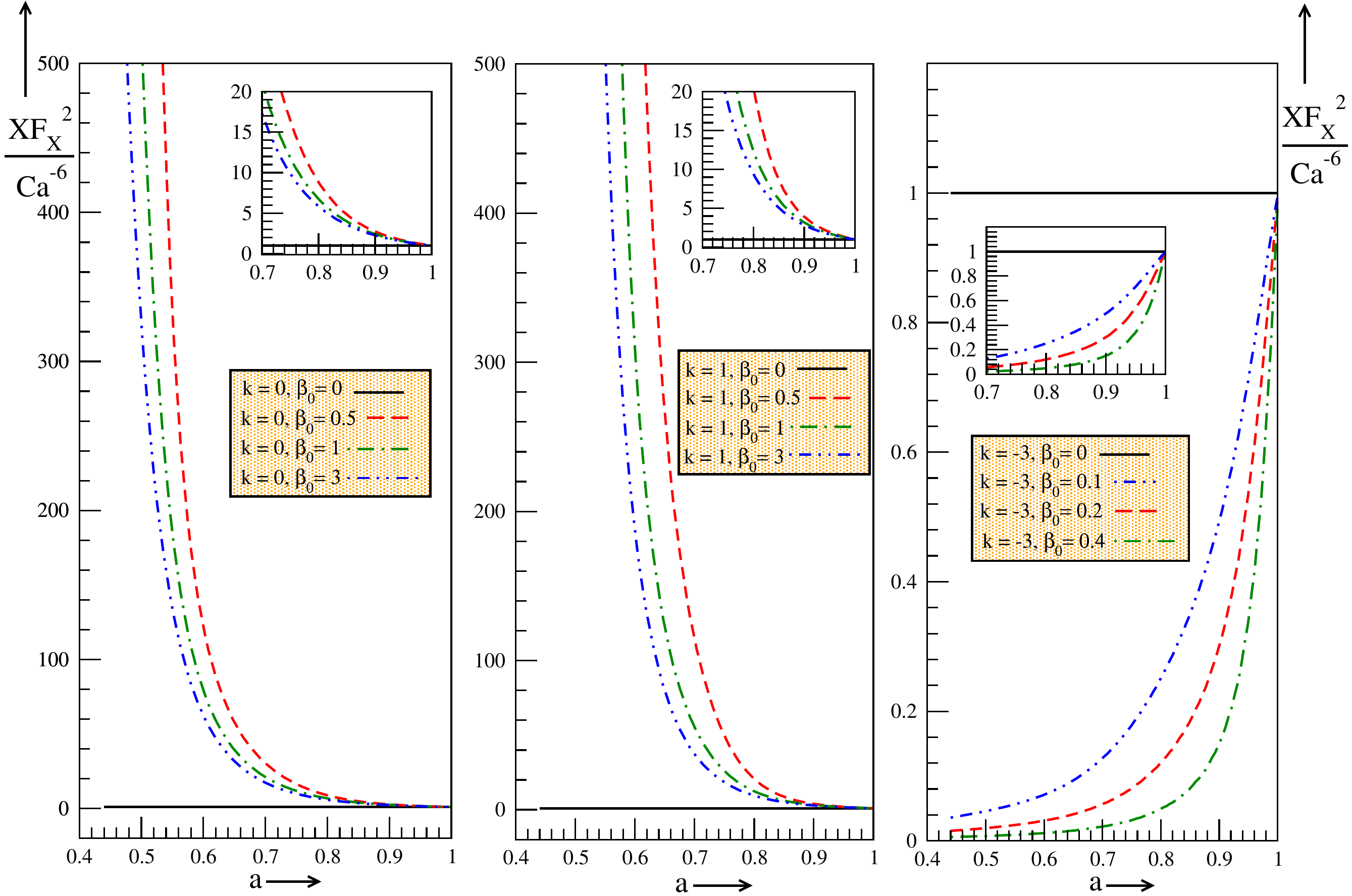} 
\caption{Plot of $XF_X^2/Ca^{-6}$ as a function
of scale factor ($a$) for some benchmark values of the model parameters $k$ and $\beta_0$. The plots within the small boxes drawn on the top of each panel depict  the same plots of the panel in an appropriately zoomed domain
 $0.7 < a < 1$  to bring out a better resolution of each of the distinct 
curves  corresponding to different $\beta_0$ values in that domain.}
\label{fig:scaling}
\end{figure}

Note that, in absence of any interaction ($\beta_0 = Q_0 / (\rho_{\rm dm}^0 + \rho_{\rm de}^0) = 0$),
the exponential term in Eq.\ (\ref{eq:sc3}) becomes unity and
the modified scaling relation  reduces to the usual form $XF_X^2 = Ca^{-6}$ as obtained in \cite{Scherrer:2004au,dePutter:2007ny}. 
Therefore, the deviation from unity, of the quantity 
$\exp\left(-2\beta_0\int_1^a\frac{ f(a_1;k)da_1}{g(a_1; k,\beta_0) } \right)$
in right hand side of
(Eq.\ (\ref{eq:sc3}) evaluated for any parameter set ($k, \beta_0$) 
gives the extent of  modification in the scaling relation due to presence
of (time-dependent) interaction between DE and DM 
parametrised in terms of ($k, \beta_0$).
The behaviour of modifications in the scaling relation are shown in three panels 
Fig. \ref{fig:scaling} for three chosen benchmark values of $k$ \textit{viz.} $k = 0$ (left panel), 
$k = 1$ (middle panel) and $k = -3$ (right panel).
In each of the panels, corresponding to a given value of $k$, we have
plotted the quantity $XF_X^2/Ca^{-6}$ as a function of the scale factor, for few
different  values of the parameter $\beta_0$ chosen
from the corresponding allowed domain of $k - \beta_0$ parameter space
discussed (and also shown in Fig.\ \ref{fig:param}) in the end of Sec.\ \ref{seciii}.
The benchmark cases: $k=0$, $k=1$ and $k=-3$
respectively correspond to $Q(t)$ = constant, $Q(t) = Q_0 a(t)$ and $Q(t) = Q_0/[a(t)]^{3}$  
in the non-conserving continuity equations\ (\ref{eq:contdm}) and  (\ref{eq:contde})
of DM and DE sectors. 
The plots shown in  Fig. \ref{fig:scaling}  show 
exponential behaviour for $\beta_0 \neq 0$ as evident from Eq.\ (\ref{eq:sc3}).
Since we have assigned the  value $(XF_X^2)\Big{|}_{a=1}$
to the constant $C$  the plots of $XF_X^2/Ca^{-6}$ 
approach  unity as the scale factor approaches to its (normalised) value $a=1$ (at present epoch).
The $\beta_0 = 0$ case, corresponds to 
non-interacting DE-DM scenario  and is represented by 
the $XF_X^2/Ca^{-6} = 1$ line in Fig. \ref{fig:scaling}.\\

\section{Conclusion}
\label{secv}
In this paper we considered 
a scenario of interacting dark matter and dark energy
during the late time evolution of cosmic evolution,
neglecting the contribution due to radiation and baryonic
matter to the total energy density. We describe the dynamics
of DE to be driven by a homogeneous $k-$essence
field ($\phi$) driven by a non-canonical Lagrangian 
of the form $L = V(\phi)F(X)$ with a constant potential
$V(\phi)=V_0$, where the dynamical term $F(X)$ is a function of
$X = \frac{1}{2}\nabla_\mu\phi\nabla^\mu\phi = \dot{\phi^2}$ 
(for homogeneous field). Under such considerations, we
showed existence of a scaling relation in the theory, which
connects the dynamical quantities $X, dF/dX$ (\textit{i.e.} $\dot{\phi}$)
to the FRW scale factor $a(t)$ of the universe along with
two relevant parameters ($\beta_0$ and $k$) of the model.
The time-dependent interaction
between DM and DE has been 
incorporated through a function $Q(t)$ in the
continuity equations ((\ref{eq:contdm}) and (\ref{eq:contde})) of the
two fluids. The source term  $Q(t)$  is parametrised
in terms of the  dimensionless parameters $\beta_0$ and $k$  
as $Q(t) = Q_0 [a(t)]^k$ where $Q_0 
\equiv \beta_0(\rho_{\rm dm}^0 + \rho_{\rm de}^0)$,
$\rho_{\rm dm}^0$ and $\rho_{\rm de}^0$ being
the present-day observed energy densities of the dark matter
and dark energy respectively. The constancy of the
$k-$essence potential is instrumental in proving
the scaling relation (Eq.\ (\ref{eq:sc3})) in interacting DE-DM scenario.
For $\beta_0=0=Q(t)$, the obtained scaling
relation reduces to the usual scaling relation
$XF_X^2 = Ca^{-6}$ obtained in \cite{Scherrer:2004au,dePutter:2007ny} 
for non-interacting DE-DM scenario.\\

We  have expressed the modification to the usual scaling relation
due to the effect of DE-DM interaction  
in terms of an exponential term of  form:
$\exp\left(-2\beta_0\int_1^a\frac{ f(a_1;k)da_1}{g(a_1; k,\beta_0) } \right)$
(see Eq.\ (\ref{eq:sc3})). In obtaining such a form we have taken
into consideration the observed feature of temporal behaviour of the FRW scale factor $a(t)$,
probed in the measurement of luminosity distance and redshift of  SNe Ia
events. This key observational ingredient enters
into the above exponential form 
at different levels of its derivation, 
%as comprehensively described in Sec.\ \ref{seciv},
through various constants 
%($D_i$'s, $A_1$'s, $P_i$'s, $R_i$'s \textit{etc.}) 
which finally got twined together  
in the obtained expressions for the
functions $f(a;k)$ and $g(a_1; k,\beta_0)$. The 
modified scaling expressed in Eq.\ (\ref{eq:sc3}),  
in the form of the exponential function,
thus encodes in it the  features of the SNe Ia data.
The values of parameters $\beta_0$ and $k$, involved
in the  scaling,
are also restricted from observed data. This has been imposed
by the condition $0 < \rho_{\rm dm}(a;k,\beta_0) < \rho(a)$, where
$\rho_{\rm dm}(a;k,\beta_0)$ 
is the dark matter density   
at an epoch corresponding to scale factor value $a$  in
the interacting DE-DM scenario and is expressed by Eq.\ ({\ref{eq:rhodmvsa}).
$\rho(a)$ is  the profile of energy density of the total dark fluid extracted 
from observation  and expressed
thorough a fitted polynomial in Eq.\ (\ref{eq:rhotot}). 
This constraint puts a bound in the $k - \beta_0$ parameter space
as shown in Fig.\ \ref{fig:param}. We also observe that, the values
of parameters $\beta_0$ and $k$ which determines the time dependence
of the source term $Q(t)$ responsible for DE-DM interactions,
decide the epoch in the past where density of DE  starts
dominating over that of DM. This has been demonstrated 
in Fig.\ \ref{fig:ET}.\\

This modified form of  scaling relation in Eq.\ (\ref{eq:sc3}),
obtained in the context of DE-DM interacting DE-DM scenario,
may also be used to eliminate $F_X$ from Eqs.\ (\ref{eq:xfx2}),
to obtain $X$ in terms of the functions 
$f(a;k)$ and $g(a_1; k,\beta_0)$,
along with the parameters $k$ and $\beta_0$. Since for
homogeneous field $\phi$, we have $X = \frac{1}{2}\dot{\phi}^2$,
one may thus obtain the scale factor dependence or
the temporal profile of the
$k-$essence scalar field in the context of DE-DM interacting scenario  using the modified scaling.\\

Finally, we also note  that, for a given set of values of the parameters  $k$ and $\beta_0$,
eliminating $F_X$ from   Eq.\ (\ref{eq:xfx1}) and the modified scaling relation (\ref{eq:sc3}),
we may  obtain the scale factor dependence of
the term $X$ as
\begin{eqnarray}
 X &=& \left[\frac{(\rho_{\rm dm}^0 + \rho_{\rm de}^0)^2}{4CV_0^2}\right]
a^6 \Big{[} g(a;k,\beta_0) \Big{]}^2 \exp\left(2\beta_0\int_1^a\frac{ f(a';k)da'}{g(a'; k,\beta_0) } \right) \,.
\label{eq:xvsa}
\end{eqnarray} 
Up to the constant multiplicative factor $ \frac{(\rho_{\rm dm}^0 + \rho_{\rm de}^0)^2}{4CV_0^2}$,
the right hand side of
the Eq.\ (\ref{eq:xvsa}) is numerically computable at different values of
the scale factor $a$, with the knowledge of
the functions   $g(a;k,\beta_0)$ and $f(a;k)$ as extracted
from observation, for any given choice of values of the parameters $\beta_0$ and $k$.
Also computing
$F(X) = p_{\rm de}(a)/V_0$  from Eq.\ (\ref{eq:kpde}), upto the factor $1/V_0$, at different
values of the scale factor $a$, we may eliminate $a$ from these two dependencies 
to obtain $V_0F(X)$ as a function of $ \frac{4CV_0^2}{(\rho_{\rm dm}^0 + \rho_{\rm de}^0)^2}  X$.
Such a dependence, as extracted from observation,  
using the modified scaling in the   constant potential $k-$essence model
contains the information in the form of $F(X)$
for given choices of the parameters $\beta_0$ and $k$ in the context
of interacting DE-DM scenario. A given choice of the constant  $C$ (appearing in scaling relation),
the value of the constant potential $V_0$ and the total  density of
dark matter and dark energy $(\rho_{\rm dm}^0 + \rho_{\rm de}^0)$ would uniquely
determine the form of the function $F(X)$.

\paragraph{Acknowledgement}\
 We thank the honourable referee for valuable suggestions. A.C. would like to thank Indian Institute of Technology, Kanpur for supporting this work by means of Institute Post-Doctoral Fellowship \textbf{(Ref.No.DF/PDF197/2020-IITK/970)}.

\paragraph{Data Availability Statement}\
No Data associated in the manuscript.

\end{document}